\documentclass[showpacs, showkeys, aps, nofootinbib]{revtex4-1}

\usepackage{amssymb,amsmath,xcolor,graphicx,xspace,colortbl,ragged2e,rotating} %
\graphicspath{{Acceleration_II_graphics/}{Acceleration_II_tcache/}{Acceleration_II_gcache/}}
\DeclareGraphicsExtensions{.pdf,.eps,.ps,.png,.jpg,.jpeg}

\begin{document}
\title{Acceleration of particles to high energy via gravitational  repulsion in the
Schwarzschild field
}
\author{Charles H. McGruder III
}
\email[E-mail me at: ]{mcgruder@wku.edu
}
\homepage[Visit: ]{}
\altaffiliation{}
\affiliation{Department of Physics and Astronomy, Western Kentucky University, Bowling Green, KY
42101
}
\date{
\today
}
\begin{abstract}Gravitational repulsion is an inherent aspect of the Schwarzschild solution of the
Einstein-Hilbert field equations of general relativity. We show that this circumstance means that it
is possible to gravitationally accelerate particles to the highest cosmic ray energies.
\end{abstract}
\maketitle

\section{Introduction
}
It is widely believed that there are only two
sources of energy available to accelerate cosmic particles to relativistic velocities - magnetic
field energy, which accelerates through magnetic connection and kinetic energy, which accelerates
through Fermi acceleration (see Drury [1] for a review). However, the recent discovery of pulsed TeV
photons from the Crab pulsar contradicts current models of relativistic cosmic
particle formation (Ansoldi et al. [2]). Here we point out that there is a third energy source -
gravitational energy, which is capable of accelerating particles to the highest cosmic ray energies
observed (
$ \sim 10^{20}$
eV).

It is well known that special relativity leads to space, time and mass dependency on
velocity. It is however, not so well known, that Einstein's theory of gravitation, general
relativity, leads to the dependency of the gravitational acceleration on velocity. In fact, in the
Schwarzschild field particles can experience gravitational repulsion as
was discovered independently by Droste [3], Hilbert [4, 5]] and Bauer [6]. After decades of debate
the existence of gravitational repulsion and the circumstances under which it occurs, was
clarified by Treder and Fritze [7] and McGruder [8]. In particular McGruder pointed out that
gravitational repulsion can only be detected by distant observers and not by observers located in
the Schwarzschild field.

We will show that gravitational repulsion leads to the acceleration of particles to
high energy. Although the concept of repulsive gravity associated with gravitating bodies may seem
strange to some, a number of authors have employed this concept [9-32].

\section{Acceleration and velocity in the Schwarzschild field
}
Einstein [33] and
independently Hilbert [34] developed field equations for general relativity (See
Janssen \& Renn [35] for details). The spherically symmetric
solution of the Einstein-Hilbert field equations for the empty space surrounding a non-rotating
point mass was discovered by Schwarzschild [36] and independently by Droste [13]:
\begin{equation}ds^{2} =\frac{dr^{2}}{1 -\frac{\alpha }{r}} +r^{2}(d\theta ^{2} +\sin (\theta )^{2}d\phi ^{2}) -(1 -\frac{\alpha }{r})dt
\end{equation}
$ds$
is the line element,
$t$
Schwarzschild time coordinate,
$r$
Schwarzschild radial coordinate,
$\theta $
colatitude,
$\phi $
longitude and
$\alpha $
is the Schwarzschild radius, which is:
\begin{equation}\alpha  =2GM
\end{equation}
where
$M$
is the mass of the gravitating body and
$c$
the speed of light is 1. When
$r \rightarrow \infty $
equation           (1) becomes the Minkowski metric of special relativity. It is
important to understand that the Schwarzschild coordinates in equation (1), as well as the
Schwarzschild velocities and accelerations are quantities measured by a distant observer, who is not
located in the gravitational field of the point mass. Thus, they are the quantities an
astronomer perceives.

 For purely radial motion the Schwarzschild
gravitational acceleration,
$\frac{d^{2}r}{dt^{2}}$
, experienced by a particle in the Schwarzschild field is (McGruder [8]):
\begin{equation}\frac{d^{2}r}{dt^{2}} =g[\frac{3}{1 -\frac{\alpha }{r}}(\frac{dr}{dt})^{2} -(1 -\frac{\alpha }{r})]
\end{equation}
where
\begin{equation}g =\frac{GM}{r^{2}}
\end{equation}
We see that it depends upon the Schwarzschild velocity,
$\frac{dr}{dt}$
. Therefore, to compute the gravitational acceleration we require an expression for the
velocity. It was first derived by Droste [3] and independently by Hilbert [4, 5] and confirmed by
Treder and Fritze [7]:
\begin{equation}(\frac{dr}{dt})^{2} =(1 -\frac{\alpha }{r})^{2} +A(1 -\frac{\alpha }{r})^{3}
\end{equation}
where A is a constant. Differentiation of equation (5) yields (3). Equation (5) reduces
for
$r \rightarrow \infty $
to:

\begin{equation}v^{2} =(\frac{dr}{dt})^{2} =1 +A
\end{equation}
where
$v$
is the velocity according to special relativity. If
$v =0$
, then equation (6) leads to
$A = -1$
. This refers to free fall motion from infinity. It is discussed in detail by Treder and
Fritze [7].

\section{Acceleration of particles to high energy         via gravitational
repulsion
}
From Droste [3] and Hilbert [4, 5] and
confirmed by Treder and Fritze [7] and McGruder [8], we learn that gravitational repulsion occurs
when:
\begin{equation}\frac{dr}{dt} >\frac{1}{\sqrt{3}}(1 -\frac{\alpha }{r})
\end{equation}
Inequality (7) tells us that all particles in the Schwarzschild field with
$\frac{dr}{dt} >\frac{1}{\sqrt{3}}$
experience gravitational repulsion. Figure 1 shows the areas of gravitational attraction
and repulsion. If the Schwarzschild field contains only a single particle, which is
moving radially outward and experiencing gravitational repulsion it will continue to do so until
$r \rightarrow \infty $
. B. Wieb VanDerMeer (private communication) has pointed out that this idea is apparently
not new because Droste [3] states (written in Dutch and translated by VanDerMeer):  ``.... the
acceleration never becomes zero; in that case there is according to equation (54) repulsion and the
velocity is maximal at infinite distance; during the motion there is always repulsion''.

As noted above when
$r \rightarrow \infty $
we have the Minkowski metric of special relativity. The special relativistic kinetic
energy is:
\begin{equation}E =(\gamma  -1)mc^{2}
\end{equation}
where
$\gamma $
is the Lorentz factor given by:
\begin{equation}\gamma  =\frac{1}{\sqrt{1 -v^{2}}}
\end{equation}
Inserting (6) into this expression we obtain:
\begin{equation}A = -\frac{1}{\gamma ^{2}}
\end{equation}
As an example if we let
$E =10^{20}$
eV in equation (8), which is approximately the highest energy cosmic rays observed and
insert the proton rest mass in (8), we obtain:
$\gamma  =1.06579$
$ \times 10^{11}$
. Plugging
$\gamma $
into equation (10) leads to:
$A = -8.80354$
$ \times 10^{ -23}$
. If we consider a single particle which is moving radially outward and obeys inequality
(7), then it will experience gravitational repulsion to
$r \rightarrow \infty $
, ending up with
$E =10^{20}$
eV. Figure 2 depicts the Schwarzschild velocity according to equation (5) out to
$r =10\alpha $
. Our numbers are only an example and it is certainly possible to achieve even higher
particle energies.

We now turn to the gravitational acceleration
in the Schwarzschild field, which is given by equation (3). Figure 3 is a plot
of equation (3) out to
$r =10\alpha $
. It shows that for our example the gravitational acceleration is always positive, meaning
that gravitational repulsion is taking place. The figure also shows that as the distance becomes
large, the acceleration approaches +2g (McGruder [8]), where g is given by equation (4).

 Finally, we note that unlike electrostatic and magnetic forces, which require
charged particles, the gravitational acceleration of elementary particles is only
slightly charge dependent as McGruder [37] has shown. This means that
uncharged particles like neutrons and neutrinos can be accelerated to very high energies too.

\begin{acknowledgments}Many thanks to Dr. and Mrs. William McCormick, whose
generous support has provided the prerequisite financial basis and most importantly the necessary
time to complete this project.
\end{acknowledgments}

\end{document}